# *HNMblock*: Blockchain technology powered Healthcare Network Model for epidemiological monitoring, medical systems security, and wellness

Naresh Kshetri
CyROC, School of Business & Technology
Emporia State University
Emporia, Kansas, USA
nkshetri@emporia.edu

Rahul Mishra
Department of Electronics & Communication
University of Allahabad
Uttar Pradesh, India
rahul3880@gmail.com

Mir Mehedi Rahman
CyROC, School of Business & Technology
Emporia State University
Emporia, Kansas, USA
mrahman2@g.emporia.edu

Tanja Steigner
School of Business & Technology
Emporia State University
Emporia, Kansas, USA
tsteigne@emporia.edu

*Abstract*— In the ever-evolving healthcare sector, the widespread adoption of Internet of Things and wearable technologies facilitates remote patient monitoring. However, the existing client/server infrastructure poses significant security and privacy challenges, necessitating strict adherence to healthcare data regulations. To combat these issues, a decentralized approach is imperative, and blockchain technology emerges as a compelling solution for strengthening Internet of Things and medical systems security. This paper introduces HNMblock, a model that elevates the realms of epidemiological monitoring, medical system security, and wellness enhancement. By harnessing the transparency and immutability inherent in blockchain, HNMblock empowers real-time, tamper-proof tracking of epidemiological data, enabling swift responses to disease outbreaks. Furthermore, it fortifies the security of medical systems through advanced cryptographic techniques and smart contracts, with a paramount focus on safeguarding patient privacy. HNMblock also fosters personalized healthcare, encouraging patient involvement and data-informed decision-making. The integration of blockchain within the healthcare domain, as exemplified by HNMblock, holds the potential to revolutionize data management, epidemiological surveillance, and wellness, as meticulously explored in this research article.

## I. INTRODUCTION

Currently, only half of the world's population has access to sufficient healthcare services [1], making the achievement of universal healthcare access an ongoing challenge. Chronic diseases have now become the leading cause of illness and disability globally, underlining the critical need for timely detection and effective treatment [2]. Despite the challenges posed by the complex healthcare environment and fragmented data storage systems [1], the incorporation of modern technology in healthcare is crucial for enhancing the efficiency and transparency of medical centers. The integration of the Internet of Things (IoT) and intelligent medical devices has enabled hospitals and other medical facilities to remotely monitor their patients, providing significant assistance to individuals managing long-term health conditions [2]. However, adding the Internet of Things (IoT) has caused security worries because of the possible flaws in data storage and transmission, especially when it comes to private patient data. Furthermore, the constraints of the current client-server architecture in IoT systems have highlighted the demand for decentralized solutions [3], prompting the healthcare industry to explore alternative frameworks and security methods. This research article presents the HNMblock framework, a Blockchain-based Healthcare Network Model that aims to facilitate epidemiological surveillance, enhance the safety of medical systems, and promote general wellness in the healthcare industry.

When it comes to providing high-quality healthcare services, one of the most significant challenges is efficient handling of patient data across the many different healthcare organizations [3]. The complexity of medical treatments frequently necessitates the safe and streamlined exchange of electronic medical records (EMRs) among numerous healthcare providers [3]. This is done to guarantee that patients receive care that is consistent and uninterrupted. This paper analyzes the potential of blockchain technology as a decentralized platform for safe and real-time access to patient records. It offers comprehensive answers to the existing barriers in healthcare data management systems. In order to solve these problems, this paper investigates the potential of blockchain technology as a decentralized platform. The study stresses how important blockchain technology is for improving data security, cyber security, cyber defense, making sure data storage is impossible to hack, and enabling smooth system-to-system operations in the healthcare field [4].

In the following sections, we explore the history and basics of blockchain in Section 2, and then dive into existing research in Section 3. Sections 4 and 5 focus on the relationship between healthcare and security, highlighting how blockchain can improve data security. Section 6 emphasizes the importance of proactive monitoring and wellness, showcasing how blockchain can support personalized healthcare. Section 7 introduces the HNMblock model, explaining its features and how it can enhance monitoring, security, and patient well-being. Finally, in Section 8, we summarize our findings and discuss the potential impact of the HNMblock model, emphasizing how blockchain can revolutionize healthcare management and promote overall patient health.

## II. BACKGROUND STUDY

In recent years, the healthcare industry has witnessed a paradigm shift in its approach to leveraging technology for enhancing patient care, ensuring data security, and monitoring public health. One of the revolutionary technologies in advancing the future at the forefront of this transformation is blockchain [5]. Blockchain, originally designed as the underlying technology for cryptocurrencies, has transcended its initial application to find innovative uses across various domains, with healthcare being a particularly promising arena [10]. This study critically assesses the integration of blockchain technology into the healthcare network model, with a specific focus on its applications in epidemiological monitoring, medical systems security, and wellness.

*1. Epidemiological Monitoring:* Effectively addressing global health challenges, particularly during pandemics, necessitates real-time epidemiological monitoring. Conventional healthcare systems encounter obstacles such as interoperability, data accuracy, and the speed of information dissemination. Blockchain, operating as a decentralized and distributed ledger technology, adeptly addresses these challenges by providing a transparent, immutable, and secure platform for recording and sharing healthcare data [9][10][11]. The inclusion of blockchain in epidemiological monitoring ensures data integrity, facilitates real-time tracking of disease spread, and encourages collaboration among healthcare providers, researchers, and public health agencies.

*2. Medical Systems Security:* In the digital age, safeguarding the security of medical systems and patient data is paramount. Ongoing threats of cybersecurity breaches and data compromises demand robust security frameworks. Blockchain introduces a resilient security paradigm through the application of cryptographic techniques, decentralized consensus mechanisms for agribusiness to the healthcare industry [6]. The immutability of blockchain records ensures tamper-proof data, establishing a robust defense against unauthorized access and manipulation. Smart contracts further elevate and automate security protocols, fostering trust among healthcare entities and patients.

*3. Wellness:* Beyond crisis management, blockchain significantly contributes to proactive and personalized healthcare strategies prioritizing wellness. The integration of wearable devices and Internet of Things (IoT) sensors connected to a blockchain network facilitates secure and transparent tracking of individual health metrics. Smart contracts automate wellness programs, incentivizing healthy behaviors and encouraging patient engagement. This empowerment enables individuals to assume control of their health, while healthcare providers can deliver personalized interventions based on accurate log-based methods and real-time data [7].

The integration of blockchain technology with healthcare networks introduces a transformative model that revolutionizes epidemiological monitoring, enhances medical systems security, and promotes individual wellness for medical wearables approved by the FDA [8]. This research explores the multifaceted impact of blockchain in the healthcare domain, underscoring its potential to reshape the landscape of healthcare delivery and management in the digital era. By unraveling the synergies between blockchain technology and healthcare objectives, this study contributes to the ongoing discourse on the transformative impact of emerging technologies in the healthcare landscape.

## III. LITERATURE REVIEW

In [9], Yang et al. (2023) proposes a novel framework utilizing consortium blockchain to revolutionize the sharing and monitoring of smart grid data. This method utilizes smart contracts for device monitoring, prepaid payments for secure transactions, and advanced encryption for data confidentiality. Based on their research, the framework successfully enhanced data audits, ensured transparency, and eliminated harmful demands. Enhanced efficiency and security in smart grid systems are achieved through the secure sharing of data among verified nodes and the use of immutable records. The consortium blockchain exhibits greater resilience against attacks due to its decentralized nature and restricted data access. The methodology guarantees efficient distribution of resources and reliable system performance even in challenging circumstances. It also effectively manages network congestion caused by fraudulent requests through rigorous verification, BLS signatures, and prepayment mechanisms.

In [10], Choudhury et al. (2019) highlights the increasing costs and challenges of multi-site clinical trials, which has led to the pursuit of innovative solutions. Traditional methods of data management have issues with data privacy, integrity, and conformity with regulations. One possible solution is the use of blockchain frameworks like Hyperledger Fabric, which encrypt data, simplify consent processes, and ensure compliance with IRB rules using smart contracts. A solution is now under development that leverages private channels, smart contracts, and a blockchain to improve data management and protocol compliance in clinical trials. Project goals include automating the process of transforming IRB-approved processes into smart contracts using ML and NLP. Expanding blockchain's application in trial management to enhance regulatory supervision and stakeholder engagement will lead to safer, more efficient, and more transparent clinical research in the future.

In [11], Kim et al. (2022) explores the potential of blockchain technology to enhance real-time monitoring systems, with a specific focus on healthcare. They emphasize the significance of data freshness in this context. The Age of

Information (AoI) is a crucial metric used to assess the timeliness of data in blockchain-enabled monitoring networks (BeMNs). The paper highlights the role of blockchain in maintaining current data and presents a new method for ensuring data freshness on permissioned blockchains like Hyperledger Fabric (HLF). The study examines the impact of blockchain characteristics, such as block size and timeout, on the occurrence of Age of Information breaches and data freshness. This study aims to enhance the management of data freshness in real-time monitoring systems and validate the influence of BeMN parameters on data freshness using both empirical and theoretical analysis. It enhances our comprehension of blockchain's capacity to enhance data dependability and promptness in critical industries, such as healthcare.

In [12], according to Kshetri (2021), Blockchain technology (BCT) provides a decentralized and revolutionary alternative to traditional centralized systems, particularly in the realm of public service provision. Through the utilization of decentralized networks and consensus mechanisms like Proof of Work and Proof of Stake, it enhances transparency and reduces the probability of corruption. Despite facing challenges related to scalability and energy consumption, ongoing research is dedicated to addressing these issues, particularly by enhancing the energy efficiency of techniques such as Proof of Stake. BCT incorporates public-key and elliptic-curve cryptography as security measures to ensure the confidentiality and authenticity of data. The integration of blockchain technology with public services may encounter technical and financial obstacles, however there are encouraging examples of its application in governance, such as the e-governance system in Estonia and the blockchain legislation in Malta, which offer strategic and economic advantages. Despite facing challenges, BCT has the capacity to transform public administration by enhancing data accessibility and promoting openness, particularly in less developed countries.

In [13], Srivastava et al. (2018) proposes a blockchain-based solution that addresses issues related to the scalability, security, and privacy of patient data in remote patient monitoring (RPM) systems. This technology has the potential to significantly transform RPM systems. In order to enhance the scalability of blockchain networks and increase transaction speeds, the study suggests implementing the GHOSTDAG protocol. Smart contracts provide automated health data administration, ensuring precision and adherence to rules, such as HIPAA. The study proposes the use of hybrid blockchain systems that combine the advantages of both public and private blockchains to enhance data accessibility and security. Remote Patient Monitoring (RPM) enhances the precision of diagnoses and treatments by integrating wearable devices with the Internet of Things (IoT). However, this advancement comes at the cost of compromising security. The proposed blockchain technique is a crucial measure for secure, efficient, and scalable healthcare data management. It prioritizes off-chain data storage to guarantee anonymity. This strategy also offers improved patient care through the implementation of new remote patient monitoring (RPM) and data exchange alternatives.

IV. HEALTHCARE AND SECURITY

In the realm of healthcare, security serves as a linchpin, safeguarding patient data, fortifying medical systems, and ensuring the trustworthiness of the entire healthcare ecosystem. The integration of blockchain technology with healthcare networks emerges as a groundbreaking solution to address the paramount concerns surrounding security within the industry.

*A. Security Foundations:*
* Data Integrity and Confidentiality: In the digital age, protecting patient data from unauthorized access, tampering, and breaches is of utmost importance. Blockchain's decentralized architecture and cryptographic principles ensure the immutability and confidentiality of medical records. Patient data, once entered into the blockchain, becomes tamper-proof, creating a secure foundation for healthcare information [14].
* Immutable Records and Trust: The immutability of blockchain records plays a pivotal role in establishing trust within the healthcare ecosystem. Every transaction, whether recording patient data or executing smart contracts, is cryptographically secured, and cannot be altered retroactively. This mitigates the risk of data manipulation and fosters trust among stakeholders and farmers [15] [16].
* Smart Contracts for Automated Security Protocols: Smart contracts in a blockchain-powered healthcare network introduce automated security protocols. These contracts enforce predefined rules and conditions, streamlining processes and minimizing the risk of human error. Automation improves efficiency and reduces the likelihood of security lapses [17].
* Decentralized Consensus Mechanisms: Blockchain's use of decentralized consensus mechanisms significantly contributes to security in healthcare. Traditional centralized systems are susceptible to single points of failure, making them attractive targets for malicious actors. Blockchain's decentralized nature distributes control and validation, reducing vulnerability to hacking or system failures [18].

*B. Healthcare and Blockchain Integration:*
* Epidemiological Monitoring: Blockchain facilitates a transparent and secure platform for epidemiological monitoring, enabling real-time tracking of disease outbreaks and data sharing among healthcare stakeholders. For instance, during a contagious disease outbreak, the decentralized and immutable nature of the blockchain allows relevant health data to be efficiently shared among authorized parties, enabling rapid response and targeted interventions.
* Medical Systems Security:

a. **Data Integrity and Interoperability:** Blockchain's decentralized ledger ensures data integrity, minimizing the risk of unauthorized access or tampering. Patient medical histories stored on the blockchain remain immutable, providing a trustworthy and accurate historical record.

b. **Smart Contracts for Access Control:** Smart contracts enable automated and secure access control to medical data.

For example, they can manage patient consent for sharing specific health information, enhancing privacy and security.
* Wellness and Preventive Healthcare: Blockchain promotes wellness by fostering a decentralized ecosystem that encourages individuals to take ownership of their health data. Users actively participating in preventive healthcare initiatives, such as sharing fitness data on a wellness application, can be rewarded through a decentralized incentive system, promoting healthier lifestyles.

In essence, the integration of blockchain technology into healthcare networks establishes a robust security framework that transcends the limitations of traditional systems. This amalgamation of technological innovation and healthcare objectives not only mitigates existing security challenges but also lays the groundwork for a more resilient and secure healthcare landscape, encompassing epidemiological monitoring, medical systems security, and wellness promotion.

## V. BLOCKCHAIN FOR HEALTH

Healthcare organizations stand to gain a great deal from utilizing blockchain technology, which could improve data security, give patients more agency, and streamline operations. It ensures better medication tracking and safe exchange of medical data. Technological, legislative, and organizational hurdles still stand in the way of this technology's widespread adoption. Despite these challenges, blockchain technology has exciting possibilities in healthcare and could lead to a better system overall.

### *5.1 Potential Applications of Blockchain Technology in the Healthcare Sector*

The implementation of blockchain technology in healthcare holds immense potential for transformative advancements. Granting individuals greater control over their personal health records fosters transparency and confidence, significantly enhancing patient autonomy. This innovation enables patients to monitor their health data and securely exchange information with their doctors, thereby enhancing collaboration and overall healthcare [18]. Blockchain technology is facilitating the consolidation of healthcare record keeping and the eradication of fragmented patient information by offering a secure and efficient system for storing medical data. This technology is crucial for effectively handling the growing number of transactions. It allows several users to access the data and ensures that the data remains accurate and reliable [18]. Moreover, the technology improves the exchange and compatibility of healthcare data. Blockchain facilitates efficient and secure data sharing by addressing interoperability challenges and prioritizing patients in the healthcare ecosystem [19]. Given the growing frequency of cybersecurity issues, blockchain technology offers robust solutions for identity management and safeguarding confidential information. It ensures the unchangeability of medical data and safeguards the confidentiality of patients' personal information [18]. Blockchain technology is essential in the pharmaceutical industry to avoid medicine counterfeiting and enhance patient safety through enhanced tracking. Moreover, blockchain technology has shown potential in managing extensive quantities of health data for public health initiatives, particularly during public health crises like the COVID-19 pandemic [19].

The wide-ranging and revolutionary potential of blockchain technology in the healthcare sector is profound [19]. It plays a critical role in maintaining patient confidentiality and facilitating predictive modeling, while also ensuring seamless communication and cooperation between healthcare facilities on a wide scale. The preservation of the consistency of health history records is crucial for precise medical interventions. Blockchain technology optimizes health insurance procedures, simplifies the secure interchange of health data, and enables the integration of artificial intelligence in healthcare, leading to enhanced diagnoses and treatment. Additionally, it assists in managing patient identities, creating new business models through safe data management, preventing security breaches in intelligent healthcare systems, and improving the authenticity and ownership of medical data. These applications collectively demonstrate the substantial influence of blockchain technology in improving data security, privacy, and operational efficiency in the healthcare sector [19].

The healthcare business is seeing a significant technological revolution as a result of the various uses of blockchain technology [20]. An in-depth analysis of blockchain's functionality uncovers valuable architectural and security methodologies, showcasing its capacity to enhance data security and healthcare efficiency [20]. This paper examines the potential applications of blockchain technology in healthcare settings [20]. It also explores seven distinct scenarios that demonstrate the versatility of blockchain in enhancing healthcare operations, particularly in terms of data security and interoperability [20]. These practical applications illustrate the versatility of blockchain technology in addressing the complex needs of modern healthcare systems. An innovative architecture has been proposed [20] to enhance the use of blockchain in healthcare and facilitate its integration across various platforms. The focus is on developing a specialized language for blockchain contracts. This unique approach, not explicitly mentioned in [21] [22], ensures exceptional security and interoperability, paving the way for a healthcare system that is more efficient, effective, real-time, and interconnected. This comprehensive approach emphasizes the transformative impact of blockchain technology using IoMT in healthcare, providing enhanced security and seamless system integration through the utilization of blockchain's complete capabilities. It offers secure, efficient, and integrated solutions, showcasing the increasing importance of blockchain in shaping the future of healthcare management and services.

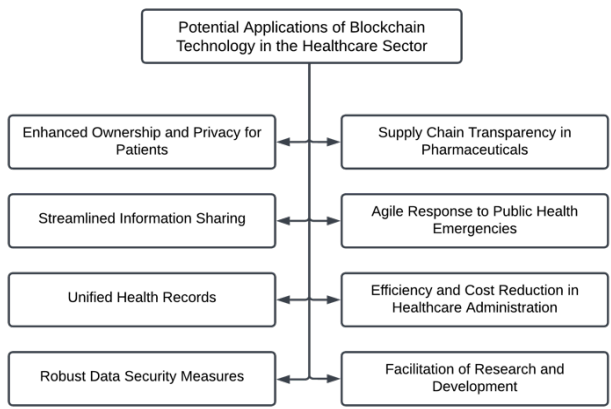

Fig 1: Potential Applications of Blockchain Technology in the Healthcare Sector

### *5.2 Challenges in Implementing Blockchain Technology in the Healthcare Sector*

Numerous impediments hinder the integration of blockchain technology in the healthcare sector. Technologically speaking, the current state of the technology is still in its nascent phase, grappling with challenges such as scalability and seamless integration with existing systems. Due to these technological hurdles, the initial expenses are higher, and the adoption processes are more complex [18]. Blockchain technology is currently not extensively employed in the healthcare sector due to organizational obstacles. Examples of these issues encompass a deficiency in trust, challenges in achieving interoperability, and struggles in demonstrating a tangible return on investment. In order to encourage the implementation of blockchain technology in healthcare organizations, it is vital to address the organizational challenges that are now taking place [18]. Privacy and regulatory issues further complicate the situation. Regulatory frameworks need to be established because of concerns arising from the decentralized nature of blockchain technology, including with data ownership, access rights, and compliance with healthcare legislation [18]. Moreover, the slow societal integration of blockchain technology in healthcare is blocked by numerous issues, namely the scarcity of effective implementations and the general lack of awareness among individuals. In order to overcome these challenges, it is imperative that education, awareness, and the development of scalable blockchain frameworks for healthcare are combined [19].

The implementation of blockchain technology in the healthcare industry encounters several obstacles, largely stemming from educational hurdles rather than technological difficulties. This necessitates healthcare professionals to adjust to new paradigms [19]. The transition from traditional healthcare models, where healthcare providers have exclusive control over patient data, to a decentralized blockchain paradigm, poses substantial challenges in terms of access control and data ownership. Significant psychological obstacles related to confidentiality, privacy, trust, and integrity are seen, in addition to practical barriers encountered by specific patient demographics, such as older adults or those with mental health conditions, in effectively managing their medical records on blockchain platforms [20]. The application of blockchain in healthcare is further complicated by the need to comply with existing privacy rules, such as HIPAA, and the challenges posed by the increasing amount of data and the resulting demands on bandwidth and storage [19].

Implementing blockchain in the healthcare business presents distinct challenges. An inherent challenge lies in the absence of universally applicable solutions, hence requiring the creation of specific strategies [20]. A common problem connected with these customized solutions is the selection of criteria that balance the need for systems to collaborate while maintaining confidence among companies and healthcare specialists [21]. Ensuring the accuracy and reliability of shared information in the blockchain poses a notable challenge, despite the well-known security and data tracking features of blockchain technology with key agreement protocol for cloud medical infrastructure [23]. In order for the implementation of blockchain technology in healthcare to achieve success, it is important to address certain existing challenges. The objective extends beyond mere technological breakthroughs, aiming to establish a system that all stakeholders can rely upon.

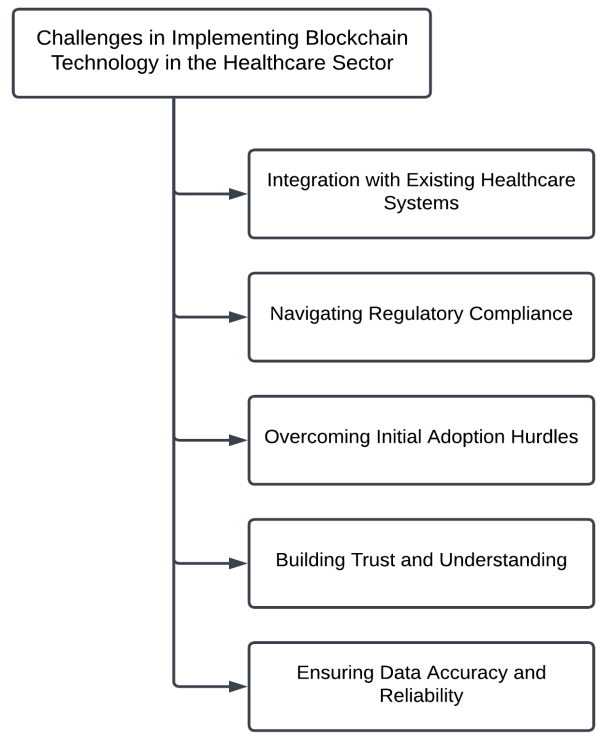

Fig 2: Challenges in implementing the Blockchain technology in the Healthcare Sector

### VI. NEED FOR MONITORING & WELLNESS

The importance of the health and well-being of humans with the use of new technologies to detect anomalies, alerts, problems, and diseases is urgent. The rise of the internet of things, embedded systems, medical supply chain, and several applications of blockchain has great potential for a range of activities in health monitoring and wellness. We use the Blockchain as the secure system, to meet constraints of medical confidentiality and information secrecy [24]. The fully automated health management system with blockchain with concerns of availability, distribution, trust, privacy has emerged in the last decade. Private blockchain has user records securely accessible to all and is a distributed ledger, concatenated database that guarantees the integrity and privacy of client records including process automation with trust, thus, bypassing middleman involvement [25] [26].

As healthcare can be revolutionized with blockchain equipped, the need for consultants from doctors, prescriptions from pharmacists, funds from insurance authorities, including hospital supplies can be possible from transactions on blockchains using the cryptocurrency [27]. Sensors in the medical devices can collect the data and can communicate with the blockchain-based application using smart contracts. As we all know that chronic disease exists over a long time, smart contracts with web3, metamask, truffle environment would support patients and doctors alert transactions [28]. Despite several known security breaches like VPNFilter, Botnet, Stuxnet, Triton, Mirai, CrashOverride etc., the sensitive treatment process of patients, blockchain model gives a very reliable and secure performance for efficiently securing healthcare applications [29].

Table 1: Blockchain-based monitoring system with some insights

| Ref. | Monitoring System | Insight 1 | Insight 2 | Insight 3 |
|---|---|---|---|---|
| [24] | IoMT-Based Platform for E-Health Monitoring | Heterogeneous embedded platform using different sensors connected via IoT to implement multi-input EHR. | More adequate by implementing the Hyperledger Fabric Blockchain, which is 100% private. | Raspberry PI is more powerful than an Arduino platform, used by pharmacists to register the medicines. |
| [25] | SHealth: health system with smart contracts | Smart-Health (SHealth), a private multi-layer blockchain-based (integrated) health management system. | All stakeholders can access health related data stored in distributed databases without authenticity. | Deploys a PBFT/PoW consensus mechanism that requires less resources & has low overhead than others. |
| [27] | HealthDote: interplanetary file system model | Blockchain-based model for health monitoring via Interplanetary File System (IFS), crypto DoteCoins. | System ensures well-being of users by data monitoring with alarm feature in emergency. | Incentive-based model that rewards patients for sharing their health data, decentralized control. |
| [28] | Smart Healthcare for Chronic-Illness Patient Monitoring | Design and prototype implementation of a blockchain system which remotely monitors patient health records. | Ethereum (difficult to tamper) is permissionless & hyper-ledger (easy to tamper) is permissioned type. | Plugins of Remix IDE, with integration of blockchain with consensus algorithm framework. |
| [29] | Care4U: Integrated Healthcare Systems | Medical sensors can transmit sensed encrypted health data via a mobile application to the doctor for privacy. | Healthcare at home using advanced technologies such as Internet of Medical Things (IoMT) but may be vulnerable. | Three connected consortium blockchain: Chain1 (images, credentials), Chain2 (data), Chain3 (process). |

## VII. PROPOSED MODEL - HNMBLOCK

The HNMblock model (proposed model) has a backbone framework of Blockchain Technology network with nodes and smart contracts. To fortify the medical security systems of several patients and their connected devices, a healthcare server supervises from the core of our proposed model as shown in Figure 1 below. The private blockchain used here for healthcare are owned by certain stakeholders or organizations (like hospitals and health centers) as compared to public blockchain. To enhance trust, smart contracts are computer programs that execute contractual obligations automatically and securely on blockchain platforms.

*A. Patient identity (in the chain or network):* The identity of patients (or the users) are verified against the immutable ledger (blockchain-enabled) for user credentials like patient's username and password.

*B. Patient(s) with private wallets/funds:* Patients are benefitted from the system model as the set of users who can access (only view) detector recommendations & own health records.

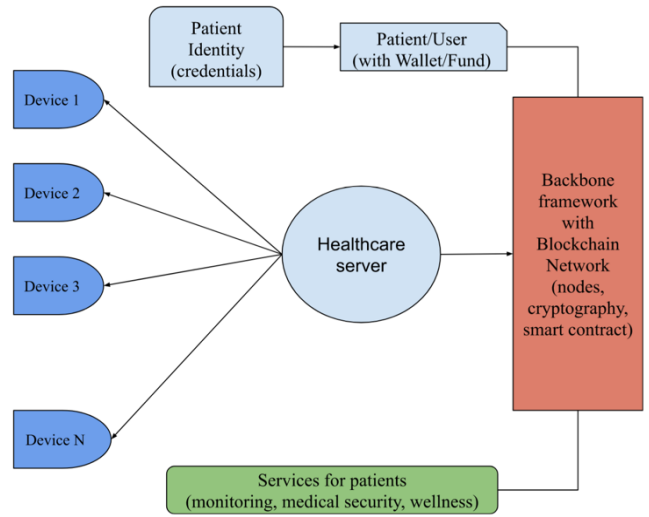

Figure 3: Proposed model - Blockchain technology (as backbone framework) powered HNMblock model with nodes, cryptography, and smart contracts

*C. Blockchain network:* The backbone framework (connected with server & services) of the HNMblock model that has nodes with advanced cryptographic techniques and smart contracts. Nodes have the previous block hash and current block hash that points to other nodes and so on.

*D. Services:* With focus on helping people / patients for timely, high-quality healthcare services, (at least for monitoring, medical systems security, and patient wellness index) the services are directly connected to the blockchain network, accessed via patient identity in the network.

*E. Healthcare Server:* The healthcare server is the centralized server that manages the connected devices (up to n devices) and acts as a bridge between the connected devices and the blockchain network that contains the blockchain nodes.

*F. Connected Devices (with medical sensors):* The devices with sensors & actuators for real-time data used by users (patients) in the HNMblock model. One patient may have more than one device connected to the healthcare server.

## VIII. Discussion & Conclusion

There are several vulnerabilities and ransomware released every hour targeting healthcare systems. Healthcare systems are vulnerable to cyberattacks due to the sensitive and valuable data they store and transmit, such as personal health information, financial data, and medical records. Cyberattacks can compromise the confidentiality, integrity, and availability (CIA) of healthcare data, as well as the safety and quality of healthcare services. The digital transformation of healthcare introduces vulnerabilities that impact patient data privacy, service delivery, and the overall integrity of healthcare systems. Today, smart healthcare systems rely on interconnected devices, electronic health records (EHRs), and telemedicine solutions. These systems are susceptible to cyber-attacks that compromise patient data, disrupt critical healthcare services, and even manipulate medical records.

Ensuring the security of these interconnected healthcare components is paramount to maintaining public trust and the confidentiality of sensitive health information. Some of the common cyber threats facing healthcare systems include replay attacks, crypto-malware, ransomware, others malware, denial-of-service (DoS), data breaches, and phishing. To prevent and mitigate these threats, healthcare systems need to adopt security standards, protocols, and technologies that can ensure data protection, privacy, and trust. Blockchain technology is one of the promising solutions that can enhance healthcare security and interoperability.

This paper introduces HNMblock, a model that elevates the realms of epidemiological monitoring, medical system security, and wellness enhancement. By harnessing the transparency and immutability inherent in blockchain, HNMblock empowers real-time, tamper-proof tracking of epidemiological data, enabling swift responses to disease outbreaks. The integration of blockchain technology (HNMblock model) with medical devices for medical systems security coupled the strategic application and emerges as a powerful paradigm. This model serves as a guide for medical partners offering blockchain nodes as the backbone framework to navigate healthcare cybersecurity. The fusion of innovative technology with our existing healthcare system will play a critical role in shaping the wellness of all patients and human beings.